\documentclass[12pt,a4paper]{article}
\usepackage{amsmath,caption}
\usepackage{longtable,threeparttable,booktabs}

\usepackage[top=1.27in, bottom=1.27in, left=1.1in, right=1.1in]{geometry}
\newcommand*{\myprime}{^{\prime}\mkern-1.2mu}

\usepackage[round]{natbib}
\usepackage{color,soul}

\DeclareCaptionStyle{italic}[justification=centering]{labelfont={bf},textfont={it},labelsep=colon}
\usepackage{graphicx,psfrag,epsf,textcomp}
\usepackage{enumerate, dsfont}
\usepackage{natbib}
\usepackage{url,xcolor}
\usepackage{booktabs, subfig, bm, paralist,mathpazo,tikz,todonotes,longtable,microtype}

\usepackage[pdftex,colorlinks=true]{hyperref}
\definecolor{darkblue}{rgb}{0,0,.6}
\hypersetup{citecolor=darkblue,linkcolor=darkblue,urlcolor=darkblue}

\newcommand{\blind}{0}

\newsavebox\CBox

\begin{document}

\def\spacingset#1{\renewcommand{\baselinestretch}{#1}\small\normalsize} \spacingset{1}

\if0\blind
{
  \title{\bf Selecting the Derivative of a Functional Covariate in Scalar-on-Function Regression}
  \author{
Giles Hooker\footnote{Postal address: Department of Statistical Sciences, 1186 Comstock Hall, Cornell University, Ithaca, NY 14853, USA; Telephone: +1 607 255 1638; Fax: +1 607 255 4698; Email: gjh27@cornell.edu}\\
Department of Statistical Sciences,
Cornell University\\
Research School of Finance, Actuarial Studies and Statistics \\
Australian National University
\vspace{0.4cm} \\
    Han Lin Shang\\
    Department of Actuarial Studies and Business Analytics \\
    Macquarie University \\
 }
  \maketitle
} \fi

\if1\blind
{
	\title{\bf Selecting the Derivative of a Functional Covariate in  Scalar-on-Function Regression}
\maketitle
} \fi

\bigskip
\begin{abstract}
This paper presents tests to formally choose between regression models using different derivatives of a functional covariate in scalar-on-function regression. We demonstrate that for linear regression, models using different derivatives can be nested within a model that includes point-impact effects at the end-points of the observed functions. Contrasts can then be employed to test the specification of different derivatives. When nonlinear regression models are defined, we apply a $J$ test to determine the statistical significance of the nonlinear structure between a functional covariate and a scalar response. The finite-sample performance of these methods is verified in simulation, and their practical application is demonstrated using a chemometric data set.

\vspace{.1in}
\noindent Keywords: model selection; variable selection; likelihood ratio test;  $J$ test
\end{abstract}

\spacingset{1.62}

\section{Introduction}

Recent advances in computer recording and storing technology facilitate the presence of functional data sets, which motivated many researchers to consider various functional regression models for estimating the relationship between predictor and response variables, where at least one variable is function-valued. The functional formulation of a linear model dates back to a discussion by \cite{HM93}, \cite{DR93}; see \cite{RS05} for a full detailed overview and \cite{RHG09} for software implementation.

Since then, models to incorporate functional variable have been extended to include generalized linear models \citep{AEV08,MM05}, additive regression  \citep{FG13,MHS+14}, polynomial models \citep{YM10}, nonparametric functional regression models \citep{FV06}, semi-functional partial linear models \citep{AV06, AV08} and many more. Because the fast development in functional regression models, it has received increasing popularity in various fields of application, such as age-specific mortality and fertility forecasting in demography \citep{HS09}, analysis of spectroscopy data in chemometrics \citep{FV02}, earthquake modeling \citep{QFV11} and ozone-level prediction \citep{QF11}.

Despite relatively mature literature on functional models, there has been little attention given to selecting which derivative of observed functional data $X(t)$ to use as a covariate. One distinguishing feature of functional data is access to multiple derivatives $X^{(k)}(t)$ of $X(t)$. It is therefore natural to consider using one or more of these as a covariate. Indeed \citet{FV02} discusses the use of semi-metrics based on derivatives for non-parametric regression, and \citet{FV09} empirically finds that the use of second derivatives provides significant performance improvement in the example data set we use below. We examine formal methods of comparing models that use different derivatives of $X(t)$.

Below we distinguish between two forms of the model. When $X(t)$ or its derivatives enter the model linearly, integration by parts provides a means of embedding smooth linear functionals of both $X(t)$ and $X^{(1)}(t)$ within a larger space:\begin{equation} \label{eq:ibp}
\int^1_0 \alpha(t) X_i^{(1)}(t) dt = \alpha(1) X_i(1) - \alpha(0) X_i(0) - \int^1_0 \alpha^{(1)}(t) X_i(t)dt.
\end{equation}
Starting from using $X_i$, we can assess whether $X_i^{(1)}$ is more appropriate by first testing the expansion of a model using $X_i(t)$ to include separate point-impact effects for the endpoints. We then formulate a contrast to test whether the reduction of the expanded model corresponds to the left-hand side of \eqref{eq:ibp}. These can both be done via $F$-tests formulated for a penalized linear regression. The same formulation allows us to reverse the inference -- to start with $X^{(1)}(t)$ and test for $X(t)$ -- and to consider changes of more than one derivative as well.

When $X^{(k)}(t)$ does not enter the model through a smooth linear operator, the formulation in \eqref{eq:ibp} cannot be applied generically. Instead, we propose a form of the $J$ test of \cite{DM81} to allow us to form nested models after estimating each separately.

\vspace{.1in}

Our paper proceeds as follows. In Section~\ref{sec:linear}, we develop linear contrasts to test the adequacy of different derivatives within a linear model specification. Although not investigated here, these methods can be readily extended to generalized linear models. Section~\ref{sec:Jtest} examines a $J$ test for explicitly nonlinear models that can also be used in conjunction with functional principal components regression. Section~\ref{sec:simulation} provides some simulation results for our methods and Section~\ref{sec:tecator} illustrates these on the Tecator data set \citep[see also][]{FV06}.

Throughout the below, we distinguish between a data generating process and a fitted model. Here we use the notation $y_i = f(X_i) + \epsilon_i$ to indicate a data generating process or a hypothesized model (where $f$ may be given by a linear model in terms of functional parameters), and $y_i \sim g(X_i)$ to indicate that we fit the model $g(X_i)$ (including any parameters) to the data. For this paper, we will only consider fitting by penalized least squares, i.e., minimizing $\sum [y_i - g(X_i)]^2 + P(g)$ where $P(g)$ is quadratic in any parameters that are used to fit. However, extensions to other likelihoods are fairly immediate.

\section{Contrasts for Tests Between Derivatives} \label{sec:linear}

This section develops formalized tests between functional derivatives used as covariates. Without loss of generality, we assume a collection of data $[y_i,X_i(\cdot)]$ for $i = 1,\ldots,n$ with each $X_i$ a function of the interval $[0,\ 1]$. We consider a model of the form
\[
y_i = \beta_0 + \int_0^1 \beta^k(t) X_i^{(k)}(t) dt + \epsilon_i, \quad \epsilon_i \sim N(0,\sigma^2),
\]
where $k$ indicates the order of the derivative to use, typically $k \in \{0,1,2\}$, and a central challenge is the choice of $k$.

A nested test can always be constructed by estimating a model that includes multiple derivatives:
\[
y_i \sim  \beta_0 + \sum_{k} \int_0^1 \beta^k(t) X_i^{(k)}(t) dt.
\]
We explore this framework below, but here we note that this test is complicated by the association between the derivatives of $X_i(t)$. These derivatives do not cover the same linear space, but their spaces do overlap considerably. In this paper, we apply integration by parts to show that to compare models with different derivatives; we can embed both models into a common space by adding a finite number of point impacts and use this to construct a set of contrasts to distinguish different derivatives.

We begin by setting up contrasts to produce tests between derivatives defined through integration by parts and then discuss the numerical implementation of tests of these contrasts within common functional data packages. We will do this for three specific tests: taking $X_i$ as a baseline and testing whether $X_i^{(1)}$ is more appropriate, the reverse procedure starting from $X_i^{(1)}$ and testing whether $X_i$ is better, and testing a change of two derivatives from $X_i$ to $X_i^{(2)}$. While these tests can be given as special cases of a more general procedure, we expect that they cover all the cases that are likely to be practically relevant.

\subsection{Taking One More Derivative}

We start by considering the first derivative as a generating model:
\begin{equation} \label{eq:xprime.mod}
y_i = \alpha + \int^1_0 \beta_1(t) X^{(1)}(t) dt  + \epsilon_i,
\end{equation}
and we may have hypothesized a model using the $0$\textsuperscript{th} derivative $X(t)$
\begin{equation} \label{eq:x.mod}
y_i = \alpha + \int^1_0 \alpha(t) X(t) dt  + \epsilon_i,
\end{equation}
and wish to test whether \eqref{eq:xprime.mod} is a more appropriate model. To carry such a test out, we need to formulate a contrast that we obtain via integration by parts in~\eqref{eq:ibp}. Here we can conduct a nested test for the adequacy of the model at $k=0$ by estimating the functional linear model augmented with point impacts at the endpoints:
\begin{equation} \label{eq:x2xprime}
y_i \sim  \alpha + \gamma_0 X_i(0) + \gamma_1 X_i(1) + \int^1_0 \gamma(t) X_i(t) dt.
\end{equation}
This model adds two degrees of freedom to the original functional linear model using just $X_i(t)$ and can thus be represented as a nested test with an appropriate contrast matrix.  However, this model is over-specified and corresponds to \eqref{eq:xprime.mod} only under the constraint:
\begin{equation} \label{eq:xprime.constr}
\gamma_0 + \gamma_1 + \int_0^1 \gamma(t) dt = 0.
\end{equation}
The left-hand side of this equation represents a linear contrast that can be tested via a Wald-type procedure or equivalently by solving for $\gamma_1$ and fitting the model
\begin{equation} \label{eq:x2xprime2}
y_i \sim \alpha + \gamma_0[X_i(0)-X_i(1)] + \int \gamma(t)[X_i(t)-X_i(1)]dt,
\end{equation}
from which it should be clear that \eqref{eq:x.mod} cannot be expressed as being nested within \eqref{eq:x2xprime2}. Because these are equivalent tests, we will take the first approach and test the agreement with model \eqref{eq:xprime.mod} through contrast.

\vspace{.3in}

We can thus define a two-stage procedure:
\begin{enumerate}
\item[1)] Test the significance of $\gamma_0$ and $\gamma_1$ in \eqref{eq:x2xprime} to assess the adequacy of using $X(t)$ as a covariate relative to the alternative $X^{(1)}(t)$.
\item[2)] Test the significance of the contrast \eqref{eq:xprime.constr} as a goodness of fit assessment of the functional linear model using $X^{(1)}(t)$ as a covariate.
\end{enumerate}
As we discuss below, the test of the contrast should, in theory, be equivalent to a comparison of \eqref{eq:xprime.mod} with \eqref{eq:x2xprime}. However, the numerical implementation of these tests in commonly-used functional data analysis software may render this correspondence inexact in practice, and we recommend assessing both the estimated \eqref{eq:xprime.mod} and \eqref{eq:x2xprime} under constraint \eqref{eq:xprime.constr} when choosing a model.

\subsection{Taking One Less Derivative}

Using similar arguments, we can also consider testing a lower-order derivative as an alternative. To illustrate this, we swap the roles of \eqref{eq:xprime.mod} and \eqref{eq:x.mod} so that we start with a model for $X^{(1)}(t)$ and consider $X(t)$ as an alternative. Here again integration by parts yields
\[
\int^1_0 \beta_0(t) X_i(t)dt = \beta_0^{(-1)}(1) X_i(1) - \beta_0^{(-1)}(0) X_i(0) - \int^1_0 \beta_0^{(-1)}(t) X_i^{(1)}(t) dt,
\]
where we have used the anti-derivative
\[
\beta_0^{(-1)}(t) = \int_{0}^t \beta_0(s) ds,
\]
and we will set $\beta_0^{(-1)}(0) = 0$ since a constant can be added arbitrarily. Thus we can test the adequacy of \eqref{eq:xprime.mod} with \eqref{eq:x.mod} as a potential alternative by fitting a model
\[
y_i \sim \delta_1 X_i(1) + \int^1_0 \delta(t) X_i^{(1)}(t) dt,
\]
and testing $H_0: \delta_1 = 0$.  As above, the additional degrees of freedom over-specify the model and an exact agreement with \eqref{eq:xprime.mod} requires the constraint:
\begin{equation*}
\delta_1 - \delta(1) = 0.
\end{equation*}
These can again be tested in a two-stage procedure.

\subsection{Moving More Than One Derivative}

The same arguments can be extended to tests moves of more than one derivative. For example, to compare 0\textsuperscript{th} and 2\textsuperscript{nd} derivatives, we can iterate integration by parts:
\begin{align*}
\int^1_0 \beta_2(t) X_i^{(2)}(t) dt & = \beta_2(1) X_i^{(1)}(1) - \beta_2(0) X_i^{(1)}(0) - {\beta_2}^{(1)}(1) X_i(1) + {\beta_2}^{(1)}(0)X_i(0)  \\ & \hspace{0.5cm} + \int^1_0 {\beta_2}^{(2)}(t) X_i(t) dt,
\end{align*}
which can be assessed by including end-point impacts for $X$ and $X^{(1)}$:
\[
y_i \sim \zeta_{00} X_i(0) + \zeta_{01} X_i(1) + \zeta_{10} X_i^{(1)}(0) + \zeta_{11} X_i^{(1)}(1) + \int^1_0 \zeta(t) X_i(t) dt,
\]
with the alternative model based on $X_i^{(2)}(t)$ corresponding to the contrasts:
\begin{align*}
\zeta_{10} + \zeta_{11} + \int^1_0 \zeta(t)dt & = 0 \\
\zeta_{00} + \zeta_{01} - \int^1_0 \left( \zeta_{10} + \int_0^t \zeta(s) ds \right) dt & = 0.
\end{align*}
The same two-step procedure can then be used to assess the fit of both models.

Beyond providing a framework for constructing nested tests to move between derivatives, the results above also allow us to understand the power that we have to distinguish between potential models. In particular, we observe from \eqref{eq:x2xprime} that these models will be indistinguishable if $\beta_1(0) = \beta_1(1) = 0$. If the $X_i(t)$ are periodic with $X_i(0)=X_i(1)$ -- if represented by Fourier components, for example -- \eqref{eq:x2xprime} is not estimable, but we also have no power if $\beta_1(0) = \beta_1(1)$. Testing in the converse direction will similarly have no power if $\int_0^1 \beta_0(t) = 0$.

\subsection{Some Numerical Comments}

We provide a numerical implementation of the test above through the penalized basis expansion framework described in \cite{RS05} and taken up in several software packages, see \textit{fda} package of \cite{RWG+18}, \textit{fda.usc} package of \cite{FO12} and \textit{refund} package of \cite{GSH+18}.  Before describing the calculation that we undertake, we first note a number of numerical issues that may make the correspondence between fitting the model \eqref{eq:x2xprime} under constraints \eqref{eq:xprime.constr} inexact and some consequences of this.

The first observation is that in many popular FDA software packages, derivatives are not necessarily represented exactly. For example, the \texttt{fda} package represents the functions $X_i$ via a basis expansion. The derivatives of such functions need not themselves be within the span of this basis, but \texttt{deriv.fd} will create $X_i^{(k)}$ as functional data by projecting the derivative onto the basis expansion for $X_i$. This introduces a numerical error into the integration by parts formula \eqref{eq:ibp} whose severity depends on the basis used and the smoothness of the $X_i$.

Additionally, any estimate for $\beta_k(t)$ is subject to bias associated with the basis expansion or the smoothing penalty. Thus in \eqref{eq:x2xprime} it may be easier, with finite data, to estimate $\gamma(t)$ with target $-{\beta_1}^{(1)}(t)$ than to estimate $\beta_1(t)$ directly in \eqref{eq:xprime.mod} or {\em vice versa}.  To account for this, we have introduced a noncentrality parameter in the tests we describe in Section~\ref{sec:implementation}. However, these biases can still affect the level or power of the test, particularly when one representation of the relationship is significantly smoother than another.


Both of these observations mean that the observed squared error from fitting \eqref{eq:x2xprime} under constraints \eqref{eq:xprime.constr} may be different from that for fitting \eqref{eq:xprime.mod} despite these being theoretically equivalent. If, as we find in the Tecator data, our contrasts both conclude that \eqref{eq:x.mod} is inadequate but \eqref{eq:xprime.mod} is not, the choice of using \eqref{eq:x2xprime} versus re-fitting \eqref{eq:xprime.mod} depends on their predictive performance and the purpose of the modeling exercise.

In this paper, we have not examined the use of functional principal components regression \citep{YMW05}. If we use the eigenfunctions for some derivative $X^{(k)}$ as a basis expansion (that we hold fixed even when examining a different derivative), the calculations below remain unchanged. However, we would expect a strong bias towards using the derivative that produced the eigenfunctions. It may be more natural to project onto a different eigenbasis for each derivative in which case the $J$ test detailed in Section~\ref{sec:Jtest} can be employed, but the change of representation from $\beta_0$ to $\beta_1$ is much harder to account for mathematically.

\subsection{Implementation} \label{sec:implementation}

All the models described above can be fit through functions in one of several software packages for the FDA. Our discussion here centers on the use of a single functional covariate, but the extension to an additional scalar or functional covariates is straightforward.

We use a basis expansion $\Phi(t) = [\phi_1(t),\ldots,\phi_K(t)]$ to represent $\beta^k(t)$ and define the design matrix
\[
\left[Z^k\right]_{ij} = \int_0^1 X^{(k)}_i(t) \phi_j(t) dt,
\]
so that $\int_0^1 X_i^{(k)}(t) \gamma(t) dt= Z^k_{i \cdot} \mathbf{g}$ for a vector of coefficients $\mathbf{g}$. We write $X_0$ and $X_1$ for the vectors containing the $X_i(0)$ and $X_i(1)$ respectively and we assume that a quadratic smoothing penalty is applied to $\gamma(t)$ that can be represented as $\mathbf{g}^{\top} P \mathbf{g}$ for some matrix $P$ \citep[e.g. see][]{RHG09}.

We then estimate parameters in \eqref{eq:x2xprime} by minimizing
\[
\left|\left| Y  - \alpha - \gamma_0 X_0 - \gamma_1 X_1 - Z^0 \mathbf{g} \right| \right|^2 + \lambda \mathbf{g}^{\top} P \mathbf{g},
\]
which gives
\[
\tilde{\mathbf{g}} = \left( \tilde{Z}^{\top} \tilde{Z} + \lambda \tilde{P} \right)^{-1} \tilde{Z}^{\top} Y,
\]
using the augmented objects $\tilde{\mathbf{g}} = (\alpha,\gamma_0,\gamma_1,\mathbf{g})$, $\tilde{Z} = [1,X_0,X_1,Z^0]$ and $\tilde{P}$ contains $P$ preceded by three rows and columns of 0's.

We can estimate $\sigma^2$ from
\[
\hat{\sigma}^2 = \frac{1}{n - \mbox{df}} \left| \left| Y - \tilde{Z} \tilde{\mathbf{g}} \right| \right|^2,  \  \mbox{df} = \mbox{tr} \left( \tilde{Z}\left( \tilde{Z}^{\top} \tilde{Z} + \lambda \tilde{P} \right)^{-1}\tilde{Z}^{\top} \right).
\]
Using the sandwich matrix
\[
V =  \left( \tilde{Z}^{\top} \tilde{Z} + \lambda \tilde{P} \right)^{-1} \tilde{Z}^{\top} Z \left( \tilde{Z}^{\top} \tilde{Z} + \lambda \tilde{P} \right)^{-1},
\]
we can obtain an $F$ statistic for the contrast $C \tilde{\mathbf{g}}$
\[
F = \frac{1}{p_C \hat{\sigma}^2} \tilde{\mathbf{g}}^{\top} C^{\top} \left( C^{\top} V C \right)^{-1} C \tilde{\mathbf{g}},
\]
where we are interested in the contrast matrices
\[
C_1 = [0_{2 \times 1} \ I_{2 \times 2} \ 0_{2 \times k}]
\]
to assess the significance of $\gamma_0$ and $\gamma_1$ and
\[
C_2 = [0 \ 1 \ 1 \ \mathbf{m}]
\]
to test \eqref{eq:xprime.constr} where $m_j = \int \phi_j(t) dt$ and $p_C$ is the dimension of the column space of $C$.

Under the null hypothesis, and ignoring smoothing and numerical biases, $F$ should be distributed as an $F$ statistic with degrees of freedom corresponding to $p_C$ and $n - \mbox{df}$. However, smoothing can generate a significant bias in favor of the alternative, compromising the level of the test. To account for this, we introduce a non-centrality parameter as follows:
\begin{enumerate}
\item[1)] Obtain fitted values $\hat{Y}_a$ and an estimate of residual variance $\hat{\sigma}^2_a$ using the alternative model \eqref{eq:x2xprime}, using the smallest smoothing parameters that allow for model identifiability; in our implementation we used $\lambda = 10^{-11}$.
%

\item[2)] Project $\hat{Y}_a$ onto the null hypothesis space of our test as follows:
    \[
    \hat{Y}_0 = \tilde{Z}(I - C^\top (CC^\top){_1} C) (\tilde{Z}^\top \tilde{Z})^{-1} \tilde{Z}^\top \hat{Y}_a.
    \]
    This first represents $\hat{Y}_a$ in terms of the coefficients in the model \eqref{eq:x2xprime}, and then projects into the null space of $C$, being equivalent in this case to setting $\gamma_0 = \gamma_1 = 0$.

\item[3)] Re-obtain coefficients from the projected $\hat{Y}_0$
    \[
        \tilde{\mathbf{g}}_0 = \left( \tilde{Z}^{\top} \tilde{Z} + \lambda \tilde{P} \right)^{-1} \tilde{Z}^{\top} \hat{Y}_0,
    \]
    and form the non-centrality parameter
    \[
        \eta = \frac{1}{\hat{\sigma}^2_a} \tilde{\mathbf{g}}_0^{\top} C^{\top} \left( C^{\top} V C \right)^{-1} C \tilde{\mathbf{g}}_0.
    \]
\end{enumerate}
We now test $F$ against the relevant quantile of an $F$-distribution with $p_C$ and $n-\text{df}$ degrees of freedom and non-centrality parameter $\eta$. The non-centrality parameter corrects the level of the test for smoothing bias; from \eqref{eq:ibp}, if $\beta_1(0) = \beta_1(1) = 0$ using either $X(t)$ or $X^{(1)}(t)$ is equivalent and a distinction between them depends on whether $\beta_1(t)$ or ${\beta_1}^{(1)}(t)$ incurs more bias. In the context of our test, the point impacts at 0 and 1 can be correlated with an over-smoothed linear functional, thereby affecting the level of the test. We base the non-centrality parameter on the orthogonal projection of an undersmoothed model onto the null hypothesis space to estimate this bias as well as possible.

Notice that $\hat{Y}_a$ and $\hat{\sigma}_a$ are calculated as part of a search over smoothing parameter values. Thus the calculation of the non-centrality parameter only requires a second penalized regression at each value of $\lambda$.


In a similar fashion, we may test \eqref{eq:x.mod} as an alternative to a null hypothesis \eqref{eq:xprime.mod} by applying the same structure as above, forming $\tilde{Z} = [1,X_1,Z^1]$ and the same contrast $C_1$ to reject \eqref{eq:xprime.mod} and then assessing
\[
C_2 = \left[ \begin{array}{rrrr} 0 & 0 & -1 &  \Phi(1) \end{array} \right]
\]
using the analogous statistic and non-centrality parameter as above.

Similarly a comparison of the 0th with the 2nd derivative as an alternative can be made by letting $X'_0 = X^{(1)}(0)$ and $X'_1 = X^{(1)}(1)$ and setting $\tilde{Z} = [1,X_0,X_1,X'_0, X'_1, Z^0]$  and assessing
\[
C_1 = [0_{4 \times 1} \ I_{4 \times 4} \ 0_{4 \times k}]
\]
and
\[
C_2 = \left[ \begin{array}{rrrrrr} 0 & 0 & 0 & 1 & 1 & \mathbf{m} \\
0 & 1 & 1 & -1 & 0 & -\bar{\mathbf{m}} \end{array} \right]
\]
with $\bar{m}_j = \int \int_0^t \phi_j(s) ds dt$. When not available analytically, we can obtain this vector by observing that
\[
\int_0^1 \int_0^t \phi_j(s) ds dt = \int_0^1 \phi_j(t)dt - \int_0^1 t \phi(t)dt
\]
by a further integration by parts argument. The expressions above can be evaluated by, for example, the function \texttt{inprod} in the \texttt{fda} package. See the code in the supplementary materials for details.


\section{$J$ Test for More General Models} \label{sec:Jtest}

An analysis using integration by parts as described above requires $X^{(k)}(t)$ to enter the model via a smooth linear operator $\int_0^1 \beta^k(t) X^{(k)}(t) dt$.  When this is not the case -- with the non-parametric regression methods of \citet{FV02}, or in the additive models in \citet{MHS+14} --  we cannot generically embed models based on $X^{(k)}(t)$ and $X^{(j)}(t)$ for $j\neq k$ within a common space and thereby allow a nested hypothesis test. Instead, we employ the $J$-test to assess non-nested models.

Specifically, we consider
\begin{align}
H_0: \ y &= m(\mathcal{X}) + \delta \label{eq:nonpara_1}\\
H_1: \ y &= s(\mathcal{X}^{\myprime}) + \delta,\label{eq:nonpara_2}
\end{align}
where $\delta$ denotes independent normally distributed error term with mean 0 and variance $\sigma^2_{\delta}$. To test the null hypothesis, we express
\begin{equation}
y = \widehat{m}(\mathcal{X}) + \theta \widehat{s}(\mathcal{X}^{\myprime}),
\end{equation}
where $\widehat{m}(\cdot)$ and $\widehat{s}(\cdot)$ are the fitted values under the null and alternative hypotheses. Effectively, the null hypothesis is
\begin{equation*}
H_0: \ \theta = 0.
\end{equation*}
We note here that a naive test of $\theta$ within a linear model will not account for the degrees of freedom used in fitting $\widehat{m}$ and $\widehat{s}$. Instead, we employ subsample-splitting \citep[see, e.g.,][]{Jarque87} to estimate $\widehat{m}$ and $\widehat{s}$ on subsample $\mathcal{S}_1$ and conduct a test for $\theta$ on $\mathcal{S}_2$.

Our procedure is summarized below:
\begin{enumerate}
\item[1)] We fit a nonparametric scalar-on-function regression, and obtained fitted values $\widehat{m}$ using $\mathcal{S}_1$.
\item[2)] We fit a nonparametric scalar-on-function regression, and obtained fitted values $\widehat{s}$ using $\mathcal{S}_1$.
\item[3)] Via a $t$ test using $\mathcal{S}_2$, we examine the statistical significance of regression coefficient associated with $\widehat{s}$.
\end{enumerate}

\section{Simulation Examples} \label{sec:simulation}

We explore the properties of the tests described above through a simulated framework. For this, we set a generative model
\begin{equation} \label{eq:dgp}
y_i = \int_0^1 \beta(t) X_i^{(1)}(t) dt  + \epsilon_i, \quad \epsilon_i \sim N(0,0.01),
\end{equation}
where we set
\[
\beta(t) = \beta_0 + 0.5 \sin(2\pi t) + 0.3 \sin(4 \pi t) + 0.1\sin(6 \pi t),
\]
and we note that when $\beta_0 = 0$, $\beta(0) = \beta(1) = 0$ and the following models also hold
\[
y_i = \int_0^1 \beta^{(1)}(t) X_i(t) dt + \epsilon_i, \quad y_i = \int_0^1 \beta^{(-1)}(t) X_i^{(2)}(t) dt + \epsilon_i,
\]
both of which we will use as null hypotheses below. In this framework, varying $\beta_0$ allows us to test power.

We generated functional covariates $X_i$ by generating random coefficients for a Fourier basis with 25 basis functions plus linear and exponential terms. Specifically
\[
X_i(t) = d_0 + d_1 \frac{(t-1/2)}{2} + \frac{d_2 e^{(t-1/2)}}{2} + \sum_{k=1}^{12} \left( f_k e^{\min(-(k-3/2), 0)} \sin( 2 \pi k t) + g_i e^{-(k-1)} \cos( 2 \pi k t) \right),
\]
where all coefficients $d_j$, $f_j$, $g_j$ are independently normally distributed, and we have included scaling factors as part of the basis. These are then projected onto an order 6 B-spline basis with 21 knots \citep[see][]{RHG09}. The projected functions are then used as the covariates when generating the $y_i$ as in \eqref{eq:dgp}.

Throughout the following, we represent coefficient functions via a basis comprising of the functions $1, t, \{\sin(2\pi k t), \cos(2 \pi k t)\}_{k=1}^{12}$. This is a Fourier basis augmented with a linear term. The linear term is necessary to represent $\beta^{(-1)}(t) = \int_{0}^t \beta(s) ds$ when $\beta_0 \neq 0$ and we have included it in all estimation procedures.   We also take $P$ to be derived from a second derivative penalty $\int \hat{\beta}^{(2)}(t)^2 dt$.

\begin{figure}[!htbp]
\centering
\includegraphics[height=5.6cm]{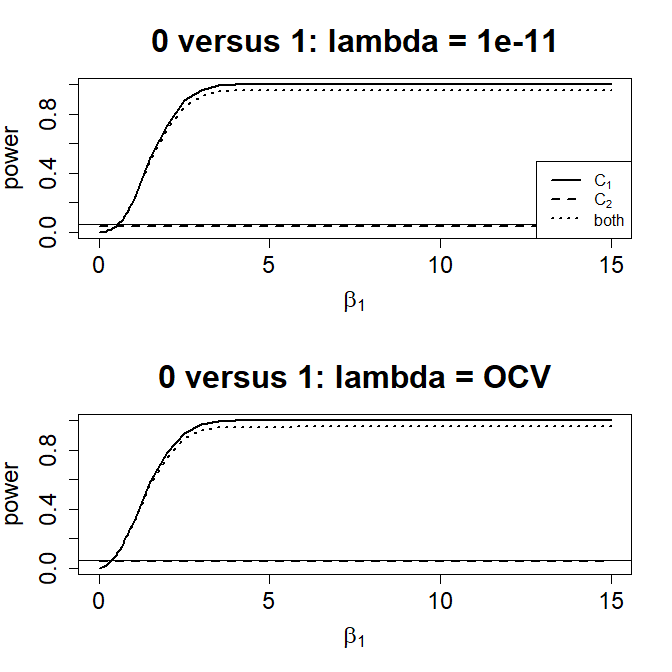} \quad
\includegraphics[height=5.6cm]{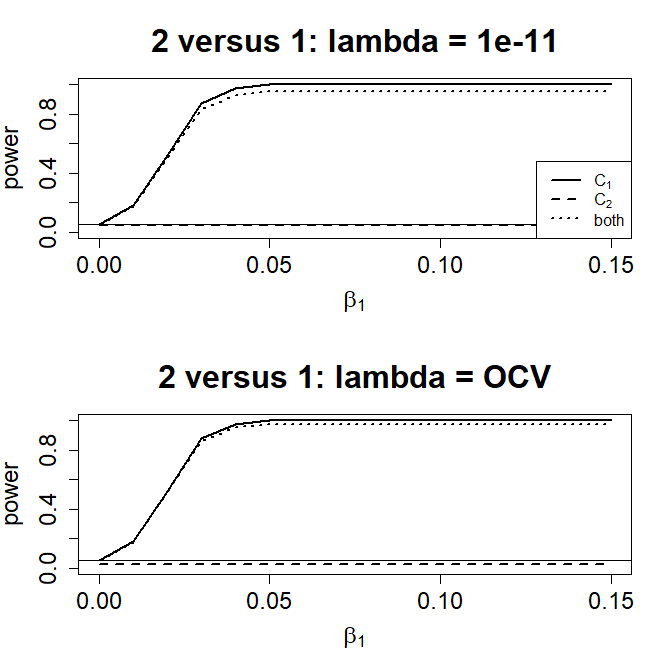} \quad
\includegraphics[height=5.6cm]{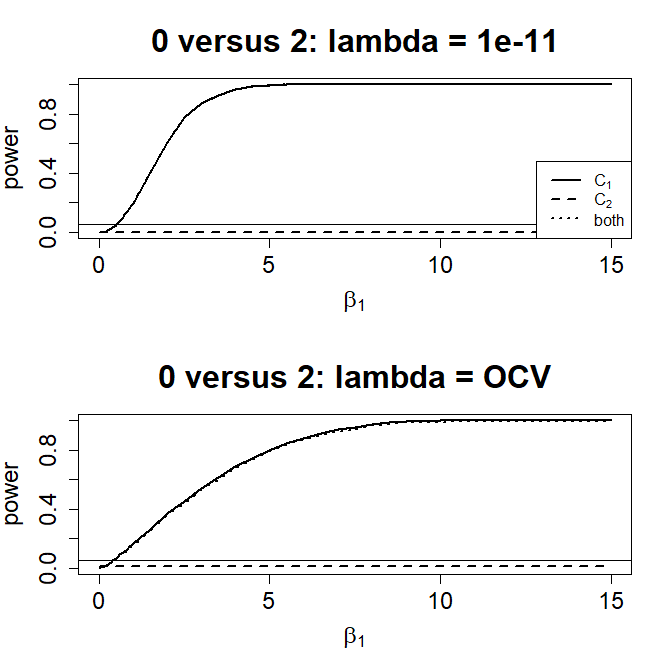} \quad
\caption{\small Power analysis of tests between different derivatives. At $\beta_1 = 0$ all models are equivalent. We observe that selecting $\lambda$ via OCV increases power, but the correlation between $X^{(2)}$ and the point impacts $(X(0), X(1))$ can result in values of $\lambda$ that compromise the level of the test in the $X$ versus $X^{(2)}$ case.} \label{fig:pow}
\end{figure}

Figure~\ref{fig:pow} demonstrates the power of the three tests detailed in Section~\ref{sec:linear} as a function of $\beta_0$. We generated a sample of 250 covariates as above and reproduced the responses $y_i$ 1000 times to obtain power. Our procedure tests two contrasts, only the first of which should be rejected, and we, therefore, do not apply a multiple testing correction. The power curves that we report give the probability of rejecting each contrast and the combined probability of arriving at the correct model: rejecting the first contrast but failing to reject the second.  We report these values both when using a small but fixed value of $\lambda$ -- enough to ensure that matrix inverses are well defined -- and for $\lambda$ chosen by ordinary cross-validation.

We find that the power of testing $X^{(2)}$ versus $X^{(1)}$ increases much more rapidly than testing $X$ versus $X^{(1)}$ (note two orders of magnitude difference in the range of the $x$-axes in Figure~\ref{fig:pow}). Using $\lambda$ chosen by cross-validation improves power relative to $\lambda=10^{-11}$ for testing $X$ versus $X^{(1)}$, but the opposite is true when testing $X$ versus $X^{(2)}$. This is likely due to smoothing effects; the relationship for $X^{(2)}$ is considerably smoother than for $X$, thus introducing significant smoothing bias and, therefore, a large non-centrality parameter. Without the non-centrality parameter, $Z$ can have a very high correlation with the point impacts $(X(0), X(1), X^{(1)}(0), X^{(1)}(1))$ which then compensate for the bias induced by smoothing penalties and resulted in rejection over 50\% of the time under the null hypothesis. By contrast, the point impacts in the test of $X^{(2)}$ versus $X^{(1)}$ are nearly uncorrelated with the functional component of the model, partly accounting for its higher power. This result suggests that including point impacts may be useful to alleviate smoothing bias in functional linear regression, whether or not they necessarily imply the use of a different derivative.

\section{Tecator Data} \label{sec:tecator}

\begin{figure}[!htbp]
\centering
\begin{tabular}{cc}
\includegraphics[width=0.48\textwidth]{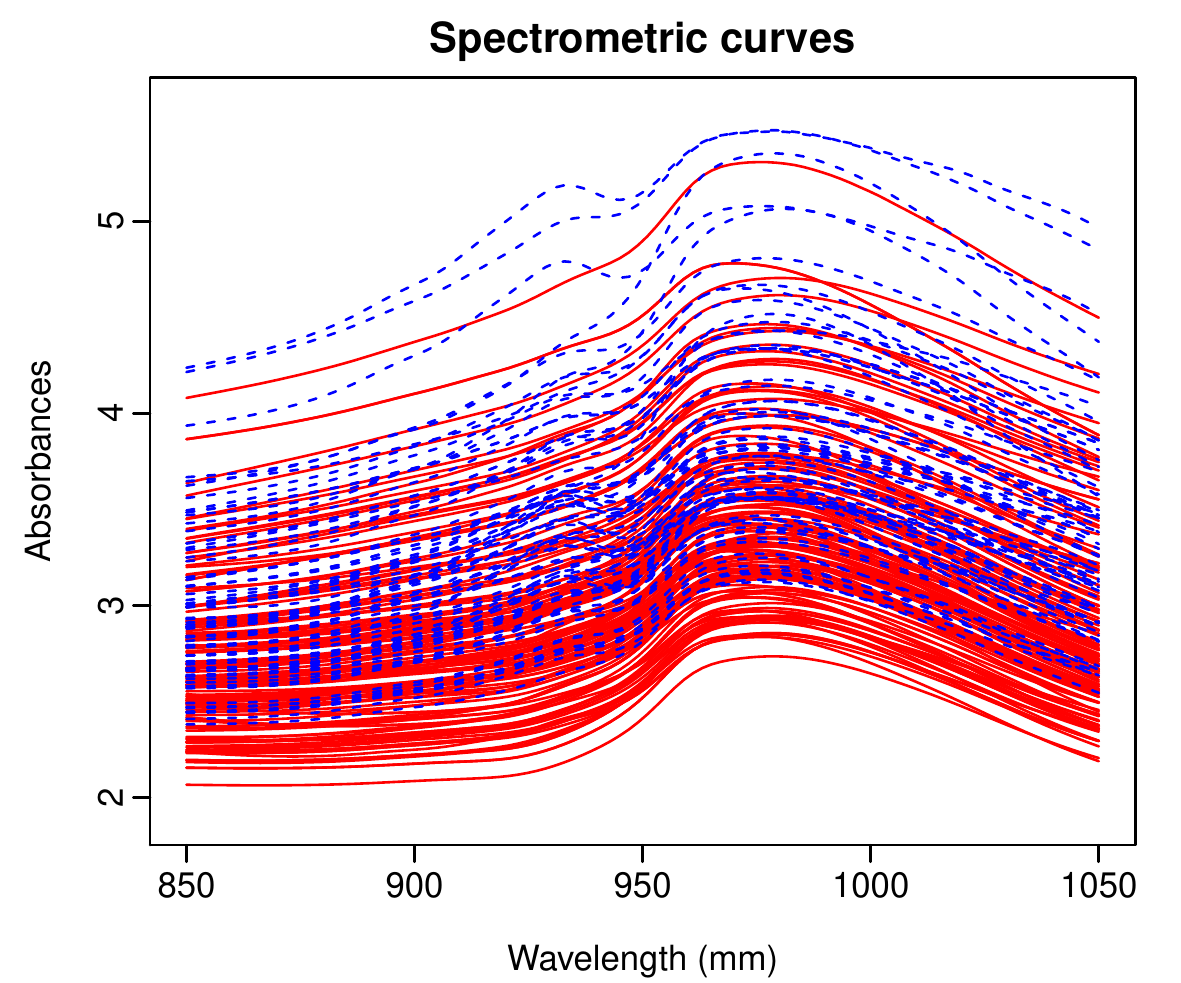} \quad
\includegraphics[width=0.48\textwidth]{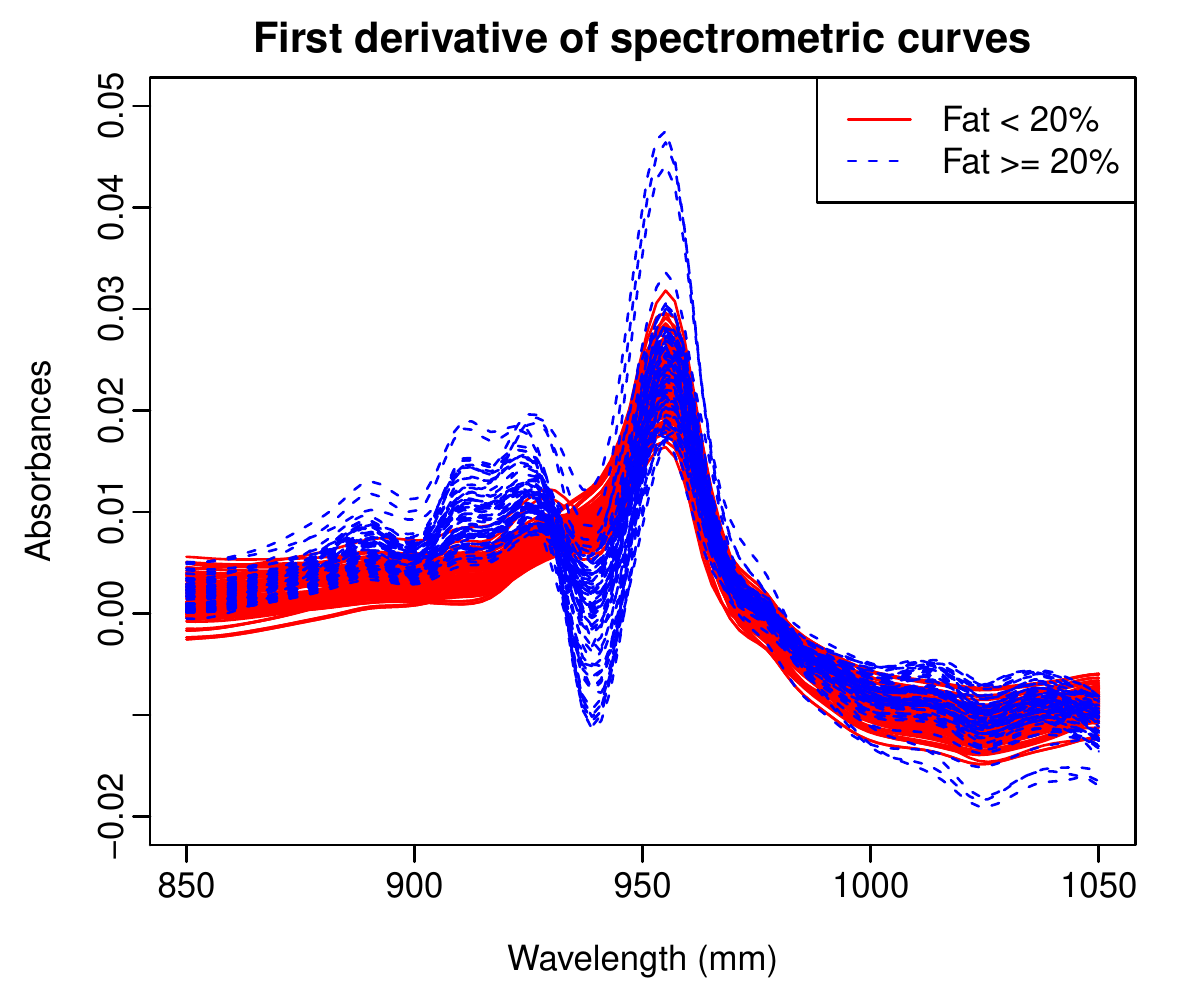} \\
\includegraphics[width=0.48\textwidth]{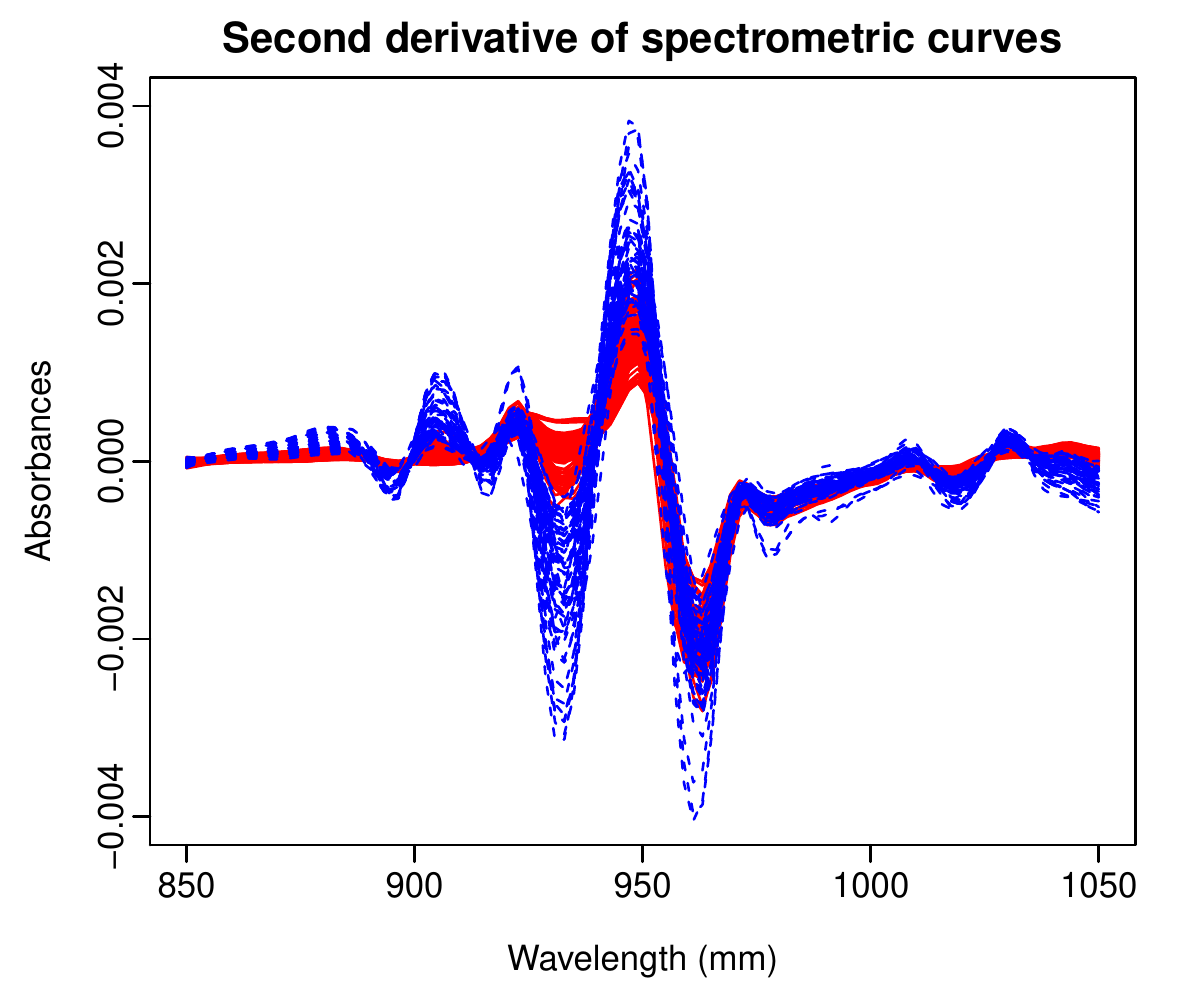} \quad
\includegraphics[width=0.48\textwidth]{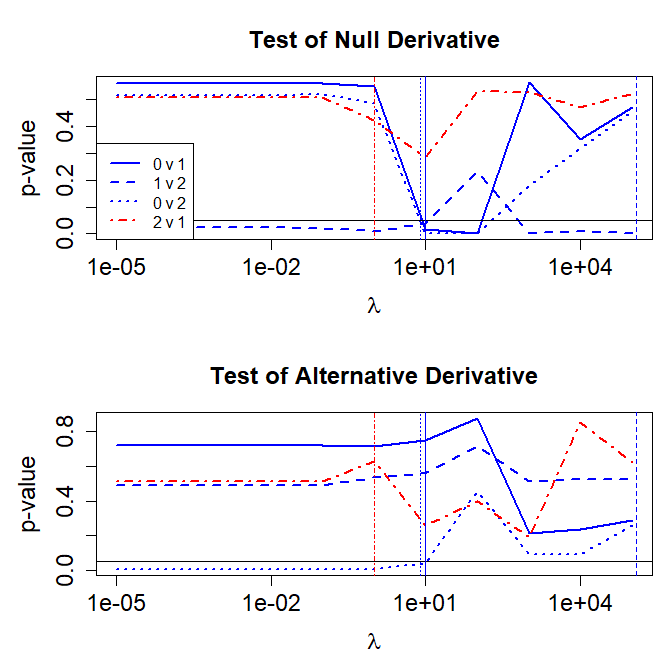}
\end{tabular}
\caption{\small Graphical displays of spectrometric curves and their 1\textsuperscript{st} and 2\textsuperscript{nd} derivatives. Curves with fat content $<20\%$ are shown in solid red lines, while curves with fat content $>=20\%$ are shown in blue dashed lines. Bottom-right gives p-values for each test between derivatives: the top panel is tests of the adequacy of the null hypothesis; the bottom panel gives p-values for the adequacy of the alternative derivative. Vertical lines indicate the cross-validated value for the test of the same line type.}\label{fig:1}
\end{figure}

We illustrate the proposed tests using an example that focuses on estimating the fat content of meat samples based on near-infrared (NIR) absorbance spectra. These data were obtained from \url{http://lib.stat.cmu.edu/datasets/tecator}, and have been studied by \cite{FV06} and \cite{AV06}, among many others. Each sample contains finely chopped pure meat with different percentages of the fat, protein, and moisture contents. For each unit $i$ (among 215 pieces of finely chopped meat), we observe one spectrometric curve, denoted by $T_i$, which corresponds to the absorbance measured at a grid of 100 wavelengths (i.e., $T_i = [T_i(t_1), \dots, T_i(t_{100})]$). We also observe its fat, protein, and moisture contents $\bm{X}\in R^3$, obtained by chemical processing. Graphical displays of the original spectrometric curves and their first and second derivatives are shown in Figure~\ref{fig:1}.

The ability to consider derivatives, as a by-product of conceiving the data as functions, is of great advantage for inference, modeling, and forecasting \citep[see, e.g.,][]{RH17} and data visualization \citep[see, e.g.,][]{Shang18}. In chemometrics, derivative spectroscopy uses first or higher derivatives of absorbance with respect to wavelength for qualitative analysis and quantification. The use of derivatives of spectral data was introduced in the 1960s, when it was shown to have many advantages \citep[see, e.g.,][]{SG64}, corroborated on this data set by the performance improvement relative to other derivatives found in \citet{FV09}.

Here, we extend the functional linear models described above to include protein and moisture content as scalar covariates and apply the procedures described in Section~\ref{sec:linear} to determine the optimal derivative use based on a linear specification:
\begin{equation}
y_i = \beta_0 + Z_i \bm{\beta} + \int \beta^k(t) X_i^{(k)}(t) dt + \epsilon_i, \label{eq:PLM}
\end{equation}
where $Z_i$ represents the linear effect of covariates, which can be incorporated naturally into the tests above. Applying our linear-specification test we find that tests of $X$ versus $X^{(1)}$, $X^{(1)}$ versus $X^{(2)}$ and $X$ versus $X^{(2)}$ all reject at the cross-validated value of $\lambda$; $X$ versus $X^{(2)}$ does reject the second test indicating possible further model elaborations, but no others do. We also tested the converse $X^{(2)}$ versus $X^{(1)}$ without rejecting. These results are displayed graphically in the bottom panel of Figure~\ref{fig:1}, where we have plotted p-values of each test as a function of $\lambda$ and indicated values minimizing cross-validation with vertical lines.


The second test for $X$ versus $X^{(2)}$ also rejects, suggesting that simply using the second derivative may not be adequate. We can assess the robustness of this conclusion by examining the same test between derivatives using the nonparametric functional regression techniques of \citet{FV02}. With the nonparametric model, we implement the $J$-test. Between the 0th and 1st derivative, we obtain a $p$-value of $<2\times 10^{-16}$, which indicates a strong preference towards the 1st derivative. Further, we compare the 1st and 2nd derivative, and we obtain a $p$-value of $<2\times 10^{-16}$, which also indicates a strong preference towards 2nd derivative. Having selected $k=2$ in~\eqref{eq:PLM}, we consider another regression model given below:
\begin{equation}
y_i = \beta_0 + Z_i \bm{\beta} + \int \beta_2(t) X_i^{(2)}(t) dt + m[X_i^{(2)}(t)] + \epsilon_i. \label{eq:PLM+nonpara}
\end{equation}
Using the $J$-test, we compare the functional partial linear model in~\eqref{eq:PLM} and a semiparametric regression model in~\eqref{eq:PLM+nonpara}. Based on the $p$-value of $5.25\times 10^{-10}$, we conclude that there may be a nonlinear effect between 2nd order derivative of the spectroscopy curve and fat content.

While the $p$-values were computed based on in-sample goodness-of-fit using all the data samples, we suggest using sample splitting to examine out-of-sample predictive accuracy. Among the 215 curves, we randomly select 160 curves as the training sample with the remaining 55 curves as the testing sample. We implement all the tests again and report the corresponding $p$-values in Table~\ref{tab:p_value}.
\begin{table}[!htbp]
\tabcolsep 0.18in
\centering
\caption{\small $p$-values for various tests for selecting optimal derivative in a functional linear model and for selecting preferable model based on either in-sample goodness-of-fit or out-of-sample predictive accuracy.}\label{tab:p_value}
\begin{tabular}{@{}llll@{}}
\toprule
& & \multicolumn{2}{c}{Criterion} \\
Model & Derivative & Goodness-of-fit & Predictive accuracy \\
\midrule
Functional linear model & 1st vs 0th & $2 \times 10^{-16}$ & $2\times 10^{-16}$ \\
& 2nd vs 1st & $2\times 10^{-16}$ & 0.929 \\
& $k$ & 2 & 1 \\
\midrule
Functional linear model vs & & & \\
Functional partial linear model  & & $5.33\times 10^{-77}$  & $1.63\times 10^{-23}$ \\
\midrule
Functional partial linear model vs & & & \\
Semiparametric model & & $5.25\times 10^{-10}$ & $7.87\times 10^{-5}$ \\
\bottomrule
\end{tabular}
\end{table}

\section{Discussion}

The use of derivatives is a feature that distinguishes functional from multivariate data. The selection of which derivative to use can make a substantial difference to the performance of functional regression. Despite the observation that derivatives can be important, there has been relatively little formal attention given to this problem.

Within a linear model, derivatives of functional covariates cover non-nested function spaces. However, we have shown that a simple integration by parts analysis allows us to embed models based on two different derivatives within a common space by adding a finite number of point impacts. This allows us to construct finite-dimensional contrasts to assess the fit of each derivative which can be tested using standard procedures. In contrast to linear models, more general models cannot be as readily embedded in a common space. Instead, we have suggested adapting the $J$ test of \cite{DM81} with subsample splitting to distinguish between two models that have already been fit.

While we have shown that these models perform well in simulation, there is clear scope for further development. An important component of our tests is a correction for the bias, since $\beta(t)$ or $\beta^{(1)}(t)$ may be easier to estimate, and this can affect the conclusions that we draw; while our non-centrality parameter appears to work well a better theoretical grounding for it would give useful guidance. Similarly, the finite-dimensional representation of $X^{(k)}(t)$ can make the implicit function theorem inexact.  In non-parametric models, we have proposed a generic framework, but this comes at the cost of sample splitting and will likely be inefficient for any given non-parametric model; more detailed analysis will need to focus on the particular model at hand.

\section*{Acknowledgments}

The first author was partially supported by NSF grants DMS-1053252 and DEB-1353039.  The second author acknowledges a sabbatical opportunity from the Research School of Finance, Actuarial Studies and Statistics at the Australian National University, and the hospitality of the Department of Statistical Science at Cornell University.

\newpage
\bibliographystyle{agsm}
\bibliography{MS-misspecification}

\end{document}